# Selective probing of longitudinal and transverse plasmon modes with electron phase-matching

Franck Aguilar,[†] Hugo Lourenço-Martins,[‡] Damián Montero,[†] Xiaoyan Li,[§]
Mathieu Kociak,[§] and Alfredo Campos[*,†]

[†]*Laboratorio Pierre y Marie Curie, Universidad Tecnológica de Panamá, 0801 Panama City, Panama*
[‡]*CEMES-CNRS, Université de Toulouse, CNRS, 31000 Toulouse, France*
[§]*Université Paris-Saclay, CNRS, Laboratoire de Physique des Solides, 91405 Orsay, France*

E-mail: alfredo.campos@utp.ac.pa
Phone: +507 67876036

## Abstract

The optical properties of metallic nanoparticles are dominated by localized surface plasmons (LSPs). Their properties only depend on the constituting material, the size and shape of the nano-object as well as its surrounding medium. In anisotropic structures, such as metallic nanorods, two families of modes generally exist, transverse and longitudinal. Their spectral and spatial overlaps usually impede their separate measurements in electron energy loss spectroscopy (EELS). In this work, we propose three different strategies enabling to overcome this difficulty and selectively probe longitudinal and transverse modes. The first strategy is numeric and relies on morphing of nano-structures, rooted in the geometrical nature of LSPs. The two other strategies exploit the relativistic and wave nature of the electrons in an EELS experiment. The first one is the phase-matching between the electron and the plasmon excitation to enhance their coupling by either tilting the sample and modifying the electron kinetic energy. The second one – polarized EELS (pEELS) – exploits the wave nature of electrons to mimic selection rules analogous to the one existing in light spectroscopies. The above-mentioned strategies are exemplified - either experimentally or numerically - on a canonical plasmonic toy model: the nano-rod. The goal of the paper is to bring together the state-of-the-art concepts of EELS for plasmonics to tackle a pedestrian problem in this field.

# Introduction

Localized surface plasmons (LSPs) are collective oscillations of free electrons confined in a metallic nanoparticle whose frequencies can be tuned by controlling the particle shape, size, and surrounding optical medium. Notably, LSPs exhibit nanoscale confinement of electric field (hot spot) with spectral response from visible to near-infrared. All these properties have enabled major developments in applications such as photothermal therapy for cancer, [1,2] biochemical sensors,[3,4] biomedical imaging[5] and information.[6]

From the literature, it is well known that the anisotropic shape of a rod sustains longitudinal and transverse plasmon modes along the longitudinal and transverse axis, respectively. In typical optical extinction experiments, the dipolar transverse mode appears at lower wavelength and exhibiting a lower cross-section than the dipolar longitudinal mode and the latter presents a more pronounced spectral shift when modifying the aspect ratio. This makes transverse modes less interesting for applications. However, it was recently proved by experiments and simulations that the intensity ratio between dipolar longitudinal and transverse modes can be modified in a rod core-shell structure, [7] even allowing the transverse mode to reach a greater cross-section than the longitudinal one which can be interesting for solar cell applications in the visible. Indeed, mostly longitudinal modes couple to light, and longitudinal ones are always at the lowest energy, typically in the infrared.[8] This highlights the importance of experimental techniques that can clearly resolve the spectral and spatial properties of both kinds of modes.

From an experimental point of view, these modes can be excited by light with near-field excitation techniques such as SNOM (scanning near-field optical microscopy)[9] or by far-field excitation techniques such as dark-field spectroscopy for single particles scattering study[10] or extinction spectroscopy for an ensemble of particles. [11] In these techniques the spatial resolution is missing or limited even for SNOM techniques. A single-particle technique to study plasmon modes with high spatial and spectral resolution is the electron energy loss spectroscopy (EELS) coupled to a scanning transmission electron microscope (STEM), which consists in a sub-nanometer electron probe impinging locally on a sample. The transmitted electrons scattered at low angles are dispersed according to their energy loss during interaction with the sample and this information is presented in a spectrum revealing the surface plasmon peaks. Additionally, at the same time, the electrons scattered at high angles are collected by an annular detector to form a high-angle annular dark-field (HAADF) image. Contrary to the case of a monochromatic plane wave of light, the electromagnetic field associated to a fast electron behaves as an evanescent wave with a wide frequency range (UV-visible-infrared) and spatially localized,[12] which excites bright and dark plasmon modes. The classical mechanism of electron energy loss in plasmonics is explained by the electric field induced in the sample which exerts a force on the incident electrons, making them to lose energy.[13] STEM-EELS technique has been used to probe the optical properties of metallic nanoparticles with unrivaled spatial resolution[14,15] and few meV spectral resolution.[16,17] STEM-EELS has also probed multipolar plasmon modes in metallic nanostructures such as nanorods,[7,8] nanotriangles,[18] nanocrosses,[19] nanosquares,[20] nanostars,[21] nanodisks,[22,23] among others. From a theoretical point of view EELS is related to optical extinction as presented in[24–26] and more generally is also related to nanooptics through the electromagnetic local density of states (EMLDOS),[27,28] furthermore, in tomographic EELS experiments a complete EMLDOS reconstruction in 3D can be made.[29] As mentioned pre-viously, a typical extinction spectroscopy experiment shows clearly dipolar longitudinal and transverse modes in nanorods and these modes are easily interpreted in terms of light polarization with respect to the rod orientation. On the other hand, in a typical EELS experiment, the information on the polarization is missing and the possibilities to excite selectively the plasmon modes are mainly limited to the impact parameter of the electron. The transverse mode in nanorods studied by EELS has shown overlapping with higher-order modes which makes interpretation of the EEL signal difficult.[30,31] A fundamental difference between optics and electron excitation techniques comes from the nature of the light that is described by vector fields (electric and magnetic) while an electron wave packet is a scalar function. However recently, a phase-shaped electron excitation was theoretically demonstrated to be an analog of light polarization by using a dipole transition vector between two (well-chosen) phase-shaped states.[32] This concept is known as polarized EELS (pEELS) where an incident broad illumination electron beam with an initial Hermite-Gauss state provides an analog to linear polarization of light. The pEELS technique can be performed to selectively excite plasmon modes as in polarized optical extinction. Herein we used both experimental and numerical tools to gain a fundamental understanding on probing longitudinal and transverse plasmon modes in nanorods by means of electron spectroscopy technique. The effect of rod aspect ratio, relativistic effects and phase-shape of electrons on the plasmonic response are studied. This work provides useful information to explore different strategies of plasmonic excitation to adress overlapping of modes by using novel tools in EELS technique.

## Results and discussion

### A. Nanorod aspect ratio study

In this section the effect of rod aspect ratio on the plasmonic response is studied. A comparison between optics and electron excitation is necessary in order to understand similarities and differences in both excitations. The relation between optical extinction and EELS is presented in a classical quasi-static approach for the case of a planewave of light and a non-penetrating focused electron beam of velocity $\vec{v} = v\hat{z}$. For a particle of any shape, a modal decomposition shows that the extinction cross section $\sigma_{ext}$ at energy $\omega$ can be represented as:[24]

$$\sigma_{ext}(\omega) = \sum_{m=dipole} A_{m,ext}\, \omega\, \Im\{-f_m(\omega)\} \tag{1}$$

where $A_{m,ext}$ is an energy independent-prefactor and $f_m(\omega)$ is a spectral function (generalized polarizability) defined as $f_m(\omega) = 2/\big((\epsilon(1+\lambda_m) + \epsilon_d(1-\lambda_m))\big)$. In this expression, $\lambda_m$ is a real eigenvalue characterizing mode m, where $\epsilon$ is the dielectric function of the metal and $\epsilon_d$ is the dielectric function of the surrounding medium. It is important to note that the sum runs only for the dipolar modes in the system (selection rule in a quasi-static regime). In a modal decomposition, the EELS probability $\Gamma(\vec{R},\omega)$ can be written as the product of the spectral function $f_m(\omega)$ by a spatial factor depending on the electron beam position:[24,26]

$$\Gamma(\vec{R},\omega) = A_{EELS} \sum_m \Im\{-f_m(\omega)\} |\phi_m(\vec{R},q)|^2 \tag{2}$$

In equation (2) the sum runs over all modes of order m and not only the dipolar ones, $A_{EELS}$ is an energy independent-prefactor and $\phi_m(\vec{R},q)$ is the Fourier transform of the plasmon eigenpotential of the order mode $m$ along the electron beam axis. The impact parameter $\vec{R}$ is the position of electron beam in the plane $(x,y)$ perpendicular to the electron beam trajectory axis and $q = \frac{\omega}{v}$. Expression (2) can also be written in term of the z component of the plasmonic electric eigenfields by using $E_m^z(\vec{R},q) = -iq\phi_m(\vec{R},q)$. Comparing equation (1) and (2), it is clear the relation between EELS and optical extinction, both signals presenting the same spectral term $\Im\{-f_m(\omega)\}$. However, the main difference is that the spectral signal of EELS has the contribution of all modes and not only the dipolar ones and these modes are modulated in intensity by their corresponding eigenpotential (or electric eigenfield) squared prefactor. Because the eigenpotential (or electric eigenfield) depends on the electron beam impact parameter, the EELS peak weights change according to the electron beam position impinging on the sample. In literature, longitudinal and transverse modes in nanorods have been modeled analytically by the Gans theory in the quasi-static regime.[33] Numerical methods have also been widely used to calculate these modes beyond the quasi-static regime, such techniques are finite-difference time-dependent (FDTD),[34] discrete dipole approxima- tion (DDA)[35] and boundary element method (BEM).[36] The latter is used in this work due to its good accuracy with experimental data and short calculation time.[37]

Gold nanorods were analyzed to investigate the influence of the aspect ratio on the plasmonic resonance. In Figure 1a and b, typical setups of optical extinction spectroscopy and electron spectroscopy are sketched, respectively. Both experimental setups present similar components, a source of light (or electron beam), optical lenses (or electromagnetic lenses), the transmitted light (or electron beam) passes through a slit (or EELS aperture) and is finally collected by a spectrometer (or magnetic prism). It is highlighted that optical extinction can be polarized. Moreover, while optical extinction studies a set of nanoparticles, electron spectroscopy studies a single nanoparticle with a sub-nanometric spatial resolution which allows focusing the impinging electron beam at different sample positions (impact parameter). Optical extinction measurements can also be performed in a UV-visible spectrophotometer in absorbance mode (Absorbance $\propto \sigma_{ext}$) for randomly oriented nanorods immersed in liquid. In this case, the light source is unpolarized and due to the different orientations of the rods, dipolar longitudinal and transverse modes are excited. In Figure 1c-e, experimental absorbance spectra of gold nanorods immersed in water are presented in which Figure 1c-d shows dipolar longitudinal and transverse modes of gold nanorods and Figure 1e presents only one peak corresponding to the limit case of a sphere (aspect ratio=1). In this limit case, the dipolar longitudinal and transverse modes converge towards the same energy position and therefore only one peak is observed. Figure 1f-i shows simulated optical extinction spectra of single gold nanorods immersed in a medium of 1.33 refractive index.

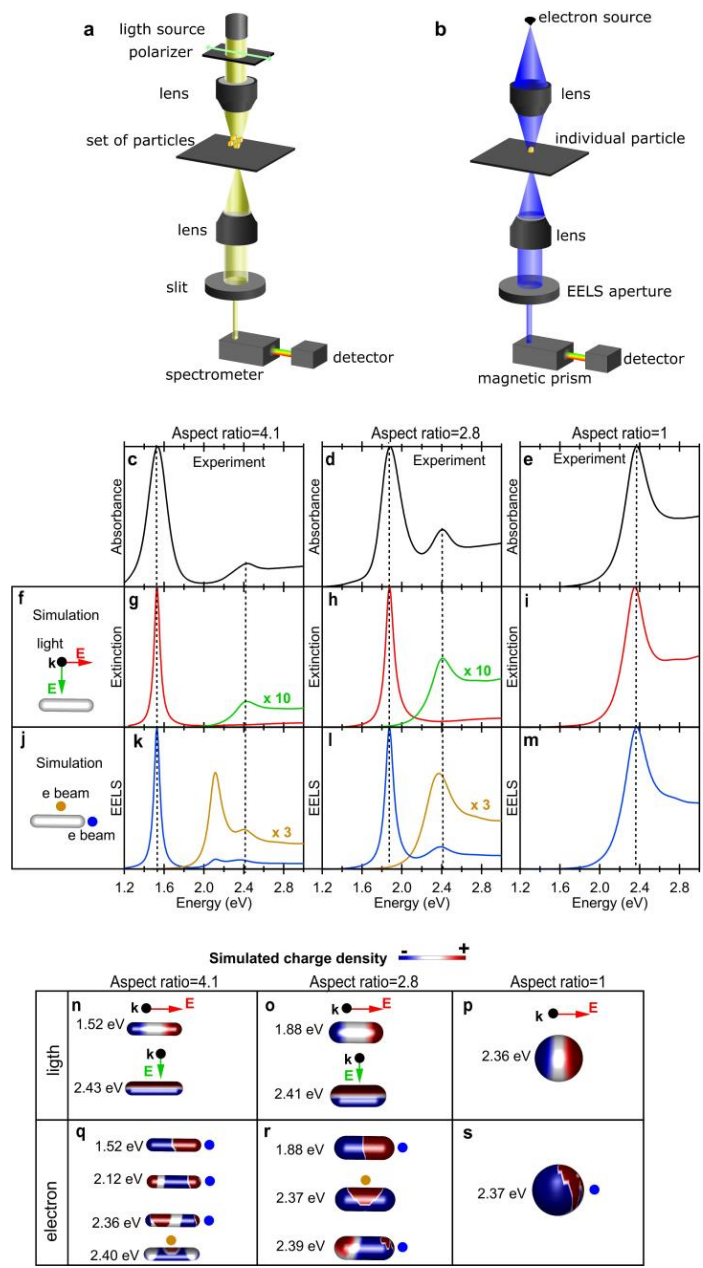

Figure 1: Comparison of optical extinction spectroscopy and electron spectroscopy. Schematic representation of typical optical extinction (a) and electron spectroscopy (b) setups. (c)-(e) Experimental absorbance spectra of gold nanorods of different aspect ratios in water. (f)-(i) Simulated optical extinction spectra of single nanorods. (j)-(m) Simulated electron spectroscopy of single nanorods using a 100 keV electron energy. (n)-(s) Charge density maps for light and electron excitation at the resonance peaks.

Simulated extinctions show identical energy positions as absorbance experiments of Figure 1c-e. The polarization of light was rotated by 90° to selectively excite the longitudinal and transverse modes (see Figure 1f). Figure 1j-m presents simulated electron spectroscopy of single gold nanorods immersed in a medium of 1.33 refractive index. An electron beam of 100 keV was used with two impact parameters, at 1 nm from the rod tip and 1 nm from the rod center, blue and brown dots in Figure 1j, respectively. The case of tip excitation clearly manifests the dipolar longitudinal mode as in absorbance experiment and simulated optical extinction. Additionally, EELS technique reveals higher-order modes that are not excited by optics (dark modes). On the other hand, the case of electron excitation at 1 nm from the rod center does not show the dipolar longitudinal mode but rather higher-order modes are obtained. To know the symmetry of the plasmon modes excited by optics and EELS, simulated charge density maps for excitation by light and electrons are plotted in Figure 1n-s. The case of light excitation clearly reveals charge maps with the symmetry of the dipolar longitudinal and transverse modes. On the other hand, for electron excitation, the dipolar longitudinal mode is well resolved as well as higher-order modes, but the dipolar transverse mode is not clear.

For further understanding of plasmonic modes excited by EELS, gold nanorods of different aspect ratios were studied experimentally in Figure 2. The energy of the electron beam was 60 keV and two electron locations were used; near the tip and near the center (see inset of Figure 2a), without touching the particle to avoid the influence of volume plasmon. The peak positions were determined by a Lorentzian fit and the EELS maps were obtained at the resonance peak by using an integration window of 0.3 eV. EEL spectra of rod aspect ratio of 1.75, show a peak at 2.13 eV and a second one at 2.43 eV. By mapping these peaks (Figure 2d), it is observed that the peak at 2.13 eV corresponds to the dipolar longitudinal mode ($L_1$) and the peak at 2.43 eV shows a high intensity around the entire particle surface that can be attributed to an accumulation of modes as will be discussed later in Figure 3. For an aspect ratio of 3.6 (Figure 2b), the tip EEL signal presents a red-shift to 1.70 eV and the center EEL signal shows a peak at 2.40 eV. The EELS plasmon maps (Figure 2e) reveal the dipolar longitudinal mode ($L_1$) for the peak at 1.70 eV and a second-order longitudinal mode ($L_2$) for the peak at 2.40 eV. By increasing the aspect ratio to 4.3 (Figure 2c), the tip EEL signal shows a red-shift to 1.46 eV and the center EEL signal a red-shift to 2.10 eV. Figure 2f reveals the maps of dipolar $L_1$ and second-order $L_2$ modes for the peaks at 1.46 eV and 2.10 eV, respectively. The evolution of plasmon energies with rod aspect ratio is presented in the energy dispersion curve of Figure 2g. The second-order longitudinal mode ($L_2$) appears for aspect ratios larger than 2.7 and at lower aspect ratios the EELS maps show a high intensity around the entire particle surface. Dipolar $L_1$ and second-order $L_2$ modes tend to red-shift with the increase of aspect ratio.

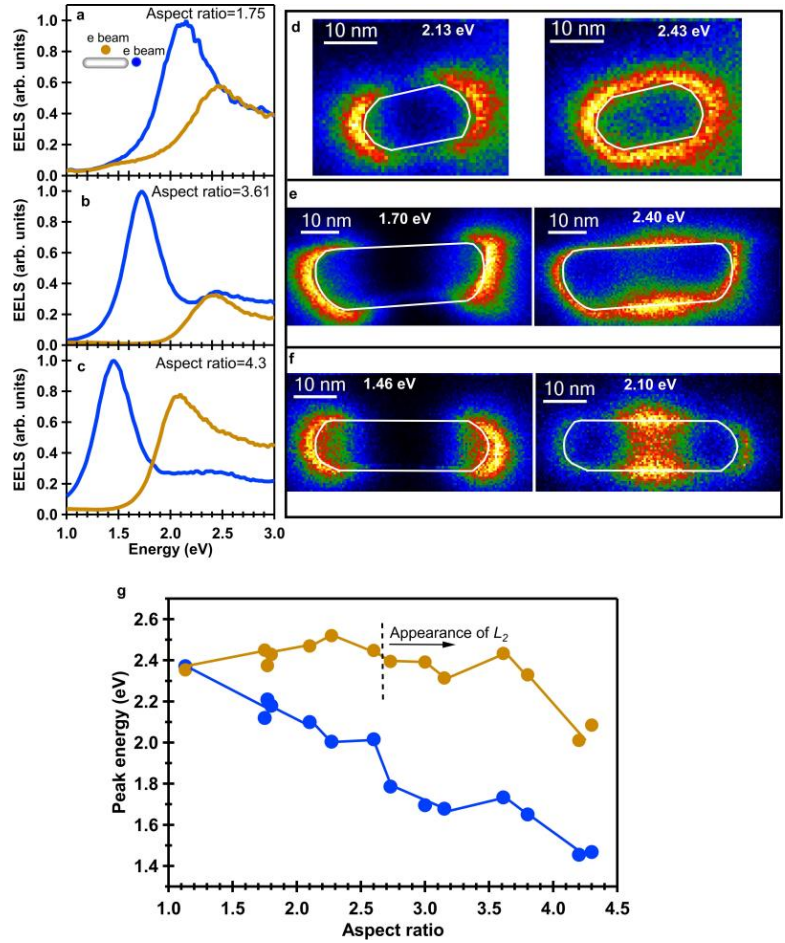

Figure 2: Experimental EEL spectra and EELS maps of gold nanorods of different aspect ratios. (a)-(c) EEL spectra at two different impact parameters: tip (blue spectra) and center (brown spectra). (d)-(f) EELS maps at energies corresponding to peaks of (a)-(c). (g) Evolution of the plasmon peak energy with the rod aspect ratio. The second-order longitudinal mode appears for rod aspect ratios larger than 2.7.

The effect of the aspect ratio on the EELS plasmonic response has also been reported for gold nanorods in a homogeneous medium.[38] The absence of a transverse plasmon mode was attributed to the overlap with the interband transitions in gold.

As presented in EEL spectra and maps of Figure 2, the dipolar longitudinal mode is well-resolved for any rod aspect ratio. However, the dipolar transverse mode is not evident. In order to understand why this mode is not well resolved in EELS, an eigenmode expansion was realized to know the order of appearances of the plasmonic peaks with the aspect ratio and investigate in depth the spectral overlapping of plasmon modes. To do this analysis we used an eigenmode expansion in a quasi-static approximation based on the BEM approach and the calculation of the spectral function $f_m(\omega)$ presented in equation (1) and (2). The eigenmode expansion shows the eigenvalues associated with the plasmon modes that can be sustained by a small nanoparticle without a specific excitation. Once the eigenvalues are known, the spectral function can be calculated. Figure 3a shows the first eigencharges symmetries of longitudinal and transverse plasmon modes in a rod. In Figure 3b, the eigenvalues $\lambda_m$ are plotted for different rod aspect ratios from 1.1 to 5, where longitudinal modes are denoted by $L_m$ and transverse modes by $T_m$ and the subscript $m$ counts for the order mode. It is observed that for an aspect ratio of 5, the first six longitudinal modes present the lowest eigenvalues followed by the dipolar transverse mode ($T_1$). On the other hand, the extreme case of an aspect ratio of 1.1 shows the dipolar longitudinal mode with the lowest eigenvalue, then the dipolar transverse mode and subsequently higher-order modes. This allows to understand that the influence of higher-order longitudinal modes on the dipolar transverse mode can be reduced by morphing a rod into a sphere. To complement this observation the imaginary of the spectral function is plotted in Figure 3c for the longitudinal modes and the dipolar transverse mode of rod aspect ratios of 1.1, 3 and 5. For the aspect ratio of 5, the dipolar transverse mode has a weak intensity and it is hidden by higher-order longitudinal modes. For an aspect ratio of 3, the dipolar transverse peak begins to be more intense but still shadowed by higher-order longitudinal modes.

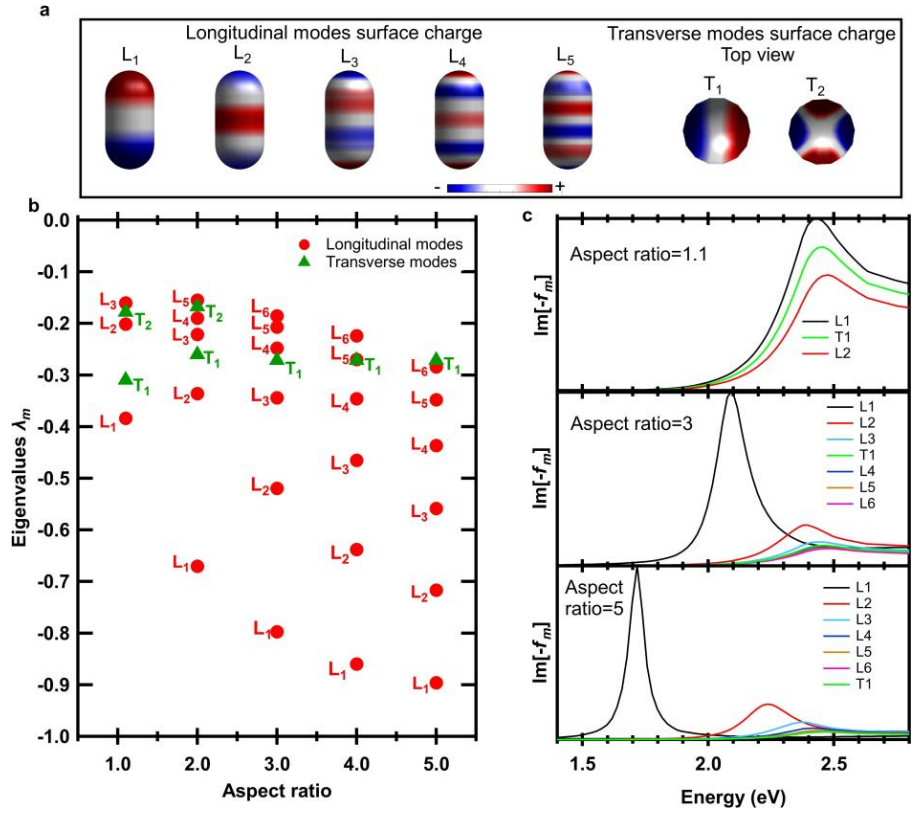

Figure 3: (a) Eigencharges calculated for a rod aspect ratio of 2. (b) Eigenvalues of different rod aspect ratios where $L_m$ denotes longitudinal modes and $T_m$ transverse modes. (c) Imaginary of the spectral function $f_m(\omega)$ calculated in vacuum for longitudinal modes and the dipolar transverse mode of gold rod aspect ratios of 1.1, 3 and 5.

For an aspect ratio of 1.1 the dipolar transverse mode is comparable in intensity to the dipolar longitudinal mode and placed almost at the same energy. In this case, the influence of higher-order longitudinal modes on the dipolar transverse mode is reduced. From Figure 3 it is now evident that a low aspect ratio can be interpreted as a rod transforming into a sphere and the dipolar longitudinal and transverse modes converging to the same energy. Due to the challenge of resolving longitudinal and transverse modes in typical EELS it is necessary to look for other strategies to solve both modes in a clear way. It is important to know that hybrids modes can appear (mix of longitudinal and transverse modes), however, they are omitted in this analysis. Furthermore, it should be noted that for spheroid-like nanoparticles, the transverse mode appears more readily as reported in Kobylko et al.[38]

## B. Electron-plasmon phase-matching: sample tilt effect

As mentioned previously, a typical EELS suffers of plasmon modes overlapping. A different strategy to explore plasmon modes is to tilt the sample with respect to the electron trajectory. This is related to the fact that EELS only measures the component of the induced electric field aligned to the electron velocity, which can be observed in the following EELS expression where both quantities are multiplied in a dot product:[13]

$$\Gamma(\omega) = \frac{e}{\pi\hbar\omega} \int dt \, \Re\{\vec{v} \cdot \vec{E}^{ind}(\vec{r}_e(t), \omega) e^{-i\omega t}\} \quad (3)$$

where $\vec{E}^{ind}$ is the induced electric field, $\vec{r}_e(t)$ is the position of the electron impinging at velocity $\vec{v}$, $e$ is the electron charge an $\hbar$ d is the reduced Planck constant. As observed in equation (3), the sample rotation allows to break the orthogonality between the induced electric field and the electron velocity, giving rise to the observation of plasmon modes not observed in a typical EELS configuration. The induced electric field $\vec{E}^{ind}$ is related to the electric field of the incoming electron $\vec{E}$ by the Maxwell equations and boundary conditions. It is well known that a slow electron exhibits a radial symmetry of the electric field, while a relativistic electron (approaching the speed of light) exhibits a flattened electric field perpendicular to its trajectory,[12,13,39] The electric field that accompanies a swift electron moving with velocity $v\hat{z}$ can be expressed as:[13,40]

$$\vec{E} = (\vec{r}, \omega) = \frac{2e\omega}{v^2 \gamma \epsilon_d} e^{\frac{i\omega z}{v}} \left[ \frac{i}{\gamma} K_0\left(\frac{\omega R}{v\gamma}\right) \hat{z} - K_1\left(\frac{\omega R}{v\gamma}\right) \hat{R} \right] \quad (4)$$

being $\gamma = 1/\sqrt{1 - \epsilon_d v^2/c^2}$ the Lorentz contraction, $c$ is the speed of light in vacuum, $\vec{r} = (\vec{R}, z)$ with $\vec{R} = (x, y)$, $K_0$ and $K_1$ are modified Bessel functions of second kind. The electron electric field can be decomposed in longitudinal ($E_z$) and transverse ($E_R$) components with respect to the electron trajectory. The $E_R$ and $E_z$ components decay exponentially at large $R$ values from the electron trajectory and for small-$R$ limit the divergence of $E_R$ ($\propto 1/R$) dominates.[13] This last effect is another reason why sample rotation is important in exploring plasmon modes. In Figure 4 the study of longitudinal and transverse modes in a gold nanorod of 2.73 aspect ratio is performed through the rod rotation both simulated and experimentally and using an electron beam of 100 keV. In Figure 4a plasmon modes excitation is studied in a gold nanorod by simulated EELS applying a rod rotation to align the rod long axis with the electron beam trajectory. The electron beam passes 5 nm from the rod surface without touching it (inset in Figure 4a). At 0°, a plasmon mode appears at 2.33 eV which is a sign of the $L_2$ mode as revealed by the charge map in the inset of Figure 4a. On the other hand, at 45° the $L_1$ mode appears at 1.85 eV and there is also a second peak at 2.40 eV, which presents similar energy to the optical extinction transverse mode $T_1$ (black dashed line). The $T_1$ mode intensity is even higher when the rod long axis is completely parallel to the beam trajectory (90°). The inset in Figure 4a presents the evolution of the induced charge with the tilting angle, which seems to indicate a transition from the second-order longitudinal mode ($L_2$) at 0° to the dipolar transverse mode ($T_1$) at 90°. Figure 4b shows the experimental EELS analysis close to the surface of a gold nanorod (see inset of Figure 4b) without rotation (0°) and with a rotation of 45°. At 0° (green curve) there is a peak centered at 2.39 eV, and at 45° (blue curve) there is an emergent peak at 1.77 eV which corresponds to the $L_1$ mode and a second peak centered at 2.41 eV that shows a slight spectral shift with respect to the non-rotated case. The complete spectral evolution of plasmon modes with the angle of rotation is detailed on Figure S1 of the Supporting Information. A complete rotation of the rod (90°) is difficult to achieve experimentally and it is important to note realizing such a configuration experimentally is rather challenging,[41] however, a rotation of ∼ 45° is enough to observe the dipolar transverse mode as presented in Figure 4a. Figure 4c shows the simulated EELS maps for rotations of 0° and 45°, where the second-order $L_2$ mode is observed at 0° and it is lost for a rotation of 45° giving rise to signs of the dipolar transverse mode ($T_1$). Figure 4d shows experimental EELS maps for rotations of 0° and 45°. Similar to simulations, the second-order $L_2$ mode is observed at 0° and signs of the dipolar transverse mode ($T_1$) appear at 45°. In order to get a closer view of what is happening when rotating the sample, the induced fields are plotted in Figure 4e and f for 0° and 90°, respectively. It is observed that at 0°, the field lines go from the rod center to the tips ($L_2$ mode), however, there are also some field lines that go to the opposite side of the rod ($T_1$ mode), which means that at 0° we excite both modes but $L_2$ dominates. For the case of 90° rotation, most of the field lines go to the opposite side of the rod ($T_1$ mode), indicating that

$T_1$ dominates.

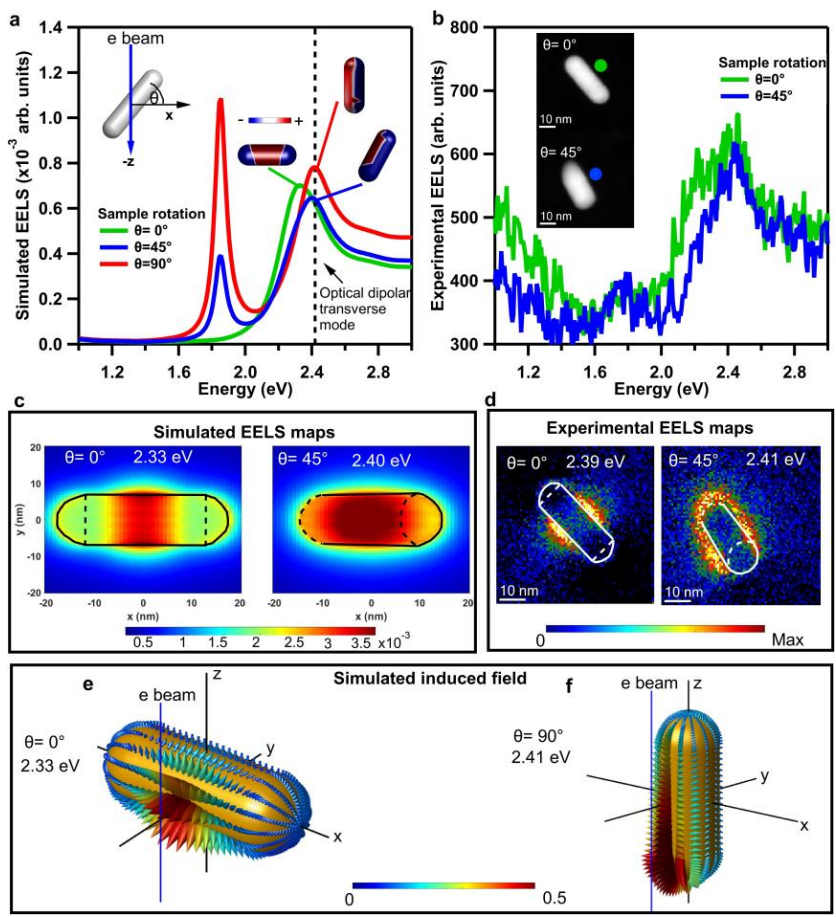

Figure 4: Rod rotation effect on EEL signal. (a) Simulated EEL spectra and charge maps for gold nanorod rotation of 0°, 45° and 90°. Refractive index of 1.33 was used as medium, the electron energy was 100 keV passing 5 nm from the rod surface and the rod aspect ratio was 2.73. Optical transverse mode energy is shown in black dashed line. (b) Experimental EEL spectra acquired close to a gold nanorod surface (color circles in the inset) for 0° and 45° rod rotation, the rod aspect ratio was 2.73 and the energy of the electron beam was 100 keV. Inset: HAADF images of a gold nanorod with rotation of 0° and 45 °. (c) Simulated EELS maps for 0° and 45° rod rotation. (d) Experimental EELS maps for 0° and 45° rod rotation.(e)-(f) Simulated induced fields at rotations of 0° and 90°.

## C. A wave mechanics approach to Electron-Plasmon phase-matching

In this section, we develop a theoretical description of EELS based on the wave nature of the electron probe. This description allows us to recover the classical expression mentioned above (equation (2)) and also extends the EELS to a quantum description that broadens the possibilities of selectivity in plasmon excitation. In this work, we aim at optimizing the coupling between the probing electron and specific plasmon modes, in order e.g., to reduce the detrimental effect of mode spectral overlap. In order to quantify this coupling, one can introduce a fundamental quantity of light-matter interaction - the electron-photon coupling constant[42] - defined as:

$$g^0(\vec{R},\omega) = \frac{e}{2\hbar\omega}\int_l dz\, E^z(\vec{R},z)\, e^{-i\frac{\omega}{v}z} \tag{5}$$

where $l$ denotes the electron trajectory and $E_z$ is the component of the induced electric field aligned with the electron trajectory. In the same way, we can introduce the *electron-photon coupling constant per mode* defined as:

$$g_m^0(\vec{R},\omega) = \frac{e}{2\hbar\omega}\int_l dz\, E_m^z(\vec{R},z)\, e^{-i\frac{\omega}{v}z} \tag{6}$$

This undimensioned complex quantity quantifies how strongly the impinging electron and a photon populated in the plasmonic mode m are coupled. Elegantly, the total EEL probability is then simply obtained by calculating:

$$\Gamma(\vec{R},\omega) = \sum_m \beta_m(\omega) \times \left|g_m^0(\vec{R},\omega)\right|^2 \tag{7}$$

Where $\beta_m(\omega) = 2(h/\pi^2)\Im\{-f_m(\omega)\}$ quantifies the spectral matching between the plasmon mode frequency and the energy-loss. Using the properties of Fourier transforms, one can trivially retrieve equation (2) from equation (7).

In order to finely analyse the effects of the electron beam properties on the electron-plasmon coupling, one needs to take into account its exact wavefunction. This can be done by employing a paraxial quantum formulation of EELS[43] i.e. by assuming that the electron beam wavefunction $\psi$ can be factorized as:

$$\psi(\vec{r}) \propto \Psi(\vec{R})e^{ikz} \tag{8}$$

where $\Psi$ represents the transverse part of the wavefunction, $e^{ikz}$ its longitudinal part and $k$ is the momentum of the electron. One can then show that the coupling constant can be generalized to arbitrary shaped electron beams:[32]

$$g_m^0(\vec{R},\omega) = \frac{e}{2\hbar\omega}\sum_f \int_{R^3} E_m^z(\vec{R},z)\Psi_i^\dagger(\vec{R})\Psi_f(\vec{R})e^{-i\frac{\omega}{v}z}d\vec{r} \tag{9}$$

where $\Psi_i$ and $\Psi_f$ represents the initial (before interaction) and final (after interaction) trans-verse electron wavefunctions. This quantity is a modal version of the vacuum electron-photon coupling constant introduced by Ofer[42] and Di Guilio et al.[44] Importantly, up to this slight modification of the coupling constant, expression (7) can be extended to the quantum regime of the probe electrons.

Interestingly, the separation in transverse and longitudinal parts inscribed in the paraxial approximation (equation (8)) translates in the expression of the coupling constant (equation (9)). Indeed, the latter is mainly composed of two integrals - the transverse overlap $I_T$ and the longitudinal overlap $I_l$ defined as:

$$I_T = \int_{R^2} E_m^z(\vec{R},z)\Psi_i^\dagger(\vec{R})\Psi_f(\vec{R})d\vec{R} \tag{10}$$

and:

$$I_l = \int_R E_m^z(\vec{R},z)e^{-i\frac{\omega}{v}z}dz \tag{11}$$

Of course, because the variables $\vec{R}$ and $z$ are coupled through $E_m^z(\vec{R},z)$, these two overlap integrals are generally intertwined and cannot be separated. In the special case of a separable fields - i.e. $E_m^z(\vec{R},z) = A_m^z(\vec{R})B_m^z(z)$ - the coupling constant can be also separated as

$$g_m^0 \propto I_T \times I_l$$

Crucially, equations (9), (10) and (11) show that the optimization of the electron-plasmon coupling boils down to the maximization of the overlap integrals by matching the symmetries of the wavefunction to the symmetry of the plasmonic mode electric field. Conventionally, the question of the optimization of $I_l$ is denoted as the longitudinal phase-matching condition (treated in section D), while the optimization of $I_T$ is called the transverse phase-matching condition (treated in section E). Even though they share the same mathematical root, these two phase-matching conditions rely on different physical processes that we will describe hereafter.

## D. Electron-Plasmon longitudinal Phase-Matching: relativistic effect

We begin by focusing on the longitudinal phase-matching condition. In order to do so, we consider a conventional point-like electron probe $\Psi(\vec{R}) = \delta(\vec{R})$. In that case, equation (9) boils down to equation (6) i.e.:

$$g_m^0 \xrightarrow[\Psi(\vec{R}) \to \delta(\vec{R})]{} \frac{e}{2\hbar\omega} I_l \tag{12}$$

In other terms, the problem of optimization of the coupling constants $g_m^0$ reduces to the optimization of the longitudinal overlap integral (11). The only free parameter in this equation is the electron speed $v$ that can be varied experimentally. To study its influence, we have simulated the EEL spectra as a function of the electron energy $E$ on Figure 5, the speed and the energy of the electron being related by the relation:

$$v = c\sqrt{1 - \frac{1}{\left(1 + \frac{E}{mc^2}\right)^2}} \tag{13}$$

Importantly, it is well known that the global EELS cross-section drops with the increasing speed of the electron. To exclude this additional effect from our analysis and focus on the phase-matching effect, we have corrected the calculation of Figure 5 by normalizing each spectrum to its total integral.

On Figure 5a and b, we calculated the EEL spectra as a function of the electron energy for an electron beam impinging along the main axis of a gold nanorod, respectively in the retarded (including all relativistic corrections) and quasi-static limits of the Maxwell equations. Interestingly, we see that, in the quasi-static limit, the longitudinal overlap integral does not depend significantly on the electron speed, while it strongly does in the retarded limit. This shows that the longitudinal phase-matching is essentially a relativistic effect. In other words, the longitudinal phase-matching is a consequence of a plasmonic field propagation effect.

This can be further understood physically by examining the trends in Figure 5a. One can notice two peaks - one at 1.85 eV and one at 2.41 eV - respectively corresponding to the above-presented dipolar longitudinal dipole mode ($L_1$) and dipolar transverse mode ($T_1$) of the rod. The dipolar longitudinal mode can be seen as an oscillation of the electron density back and forth along the main rod axis at an angular frequency of 1.85 eV/$\hbar$. The phase-matching condition consists in having the probing electron traveling along the shaft of the rod at the same speed as the charge oscillations in the rod. It is equivalent to match the speed of the traveling probe (electron beam) with the oscillation of the field (plasmon), so that the electron always sees the same electric phase in its reference frame during its transit in the interaction region. In that configuration, the plasmon wave and the electron are co-propagating at the same speed. Thus one can see that if the electron is too fast, the plasmon field will not have time to make a full cycle during the transit of the electron and will result in a positive albeit weak coupling constant. This is evidenced in the weak $L_1/T_1$ ratio at high electron energy in Figure 5a. On the contrary, if the electron is too slow, the plasmon field will oscillate several times during its transit in the near-field, so that the electric field phase will average out and result in a null coupling constant (See Figure S2 on the Supporting Information for electron energies below 1 keV). This last effect is shown in the low $L_1$ mode intensity at low electron energy (1 keV) in the inset of Figure 5a.

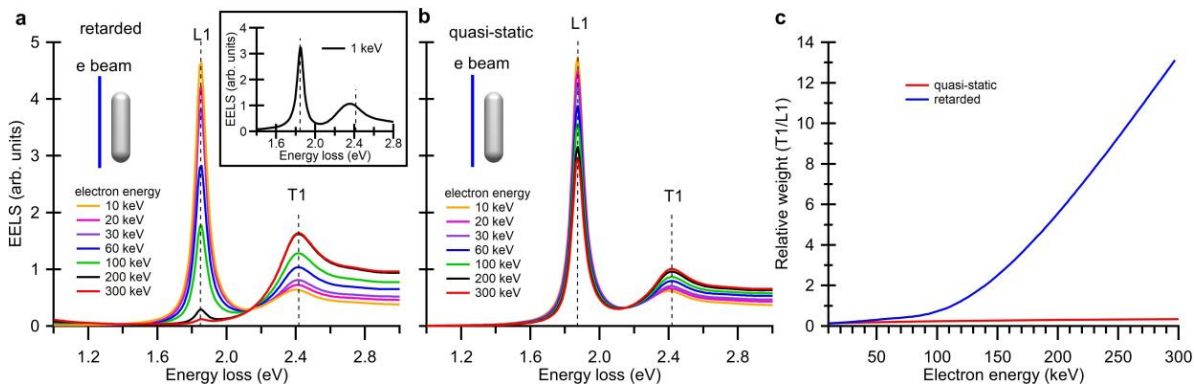

Figure 5: Effect of the electron energy on the EEL signal. The BEM simulation has been performed on a 2.73 gold nanorod aspect ratio immersed in a 1.33 refractive index medium with an electron beam impinging along the main axis of the rod (see sketch in inset), 5 nm away from the surface by (a) including field retardation effect and (b) excluding field retardation effects. (c) Intensity ratio between dipolar longitudinal ($L_1$) and transverse ($T_1$) plasmon modes for quasi-static and retarded regimes.

Now we focus on the $T_1$ mode in Figure 5a. In this case the plasmon field oscillates along the small (transverse) axis of the rod at angular frequency 2.41 eV/$\hbar$. Since now the electrons are propagating in the direction orthogonal from the field oscillation, there is not effect of co-propagation. Instead, the only requirement is that the electron travels sufficiently fast tosee a unique phase of the electric field during its transit. For too slow electrons the plasmon phase experienced by the electron will oscillate and average out during its transit, resulting in a vanishing coupling constant as presented in the inset of Figure 5a where the $T_1$ mode is lost and gives rise to longitudinal modes at lower energy position. The intensity ratio betweenthe $T_1$ and the $L_1$ mode is presented in Figure 5c. In the quasi-static regime $L_1$ dominates with ratios below 1, while in the retarded case the ratio is higher than 1, indicating that $T_1$ dominates for high electron energy and the transverse mode is selectively excited.

## E. Electron-Plasmon transverse Phase-Matching: Phase-shaped EELS

We now turn to the transverse phase-matching condition. As sketched in Figure 1a and b, unlike optical spectroscopy, conventional EELS has no polarization device. This implies a disadvantage to selectively excite plasmon modes or resolve degenerate modes in complex systems. Nevertheless, this problem has been overcome by taking advantage of the wave nature of electrons. As recently demonstrated, a phase-shaped electron beam provides an analogue of optical polarization.[32] Previous works have been demonstrated experimentally the coupling of structured electron wavefunction with plasmon potentials.[45] A description of polarized EELS (pEELS) signal in a quasi-static regime is obtained by using equation (7) and equation (9):[45]

$$\Gamma(\omega) = \frac{2e^2}{hv^2} \sum_m \sum_f \Im\{-f_m(\omega)\} \times \left| \int \Psi_f(\vec{R}) \phi_m\left(\vec{R}, \frac{\omega}{v}\right) \Psi_i^\dagger(\vec{R}) \, d\vec{R} \right|^2 \qquad (14)$$

The first sum in equation (14) is over the modes order and the second sum is over the wavefunctions final states. Expression (14) reveals a complex interference pattern. A simplified case is obtained when electrons in the in-axis direction are collected, this leads to $\Psi_f(\vec{R}) = constant$ and the pEELS depends only on the initial wavefuntion-eigenpotential coupling. In this case, the pEELS simply reads:[46]

$$\Gamma(\omega) \approx \frac{e^2 k_{max}^2}{4\pi^2 \hbar v^2 \Gamma(2)} \sum_m \Im\{-f_m(\omega)\} \times \left| \int \phi_m\left(\vec{R}, \frac{\omega}{v}\right) \Psi_i^\dagger(\vec{R}) \, d\vec{R} \right|^2 \qquad (15)$$

where $\Gamma(2)$ is the gamma function and $k_{max}$ represents the maximum out-axis wavevector collected by the aperture in the spectrometer entrance. The collection angle must be small to get $k_{max} \to 0$ and achieve a plane wavefunction as final state which is the case in equation (15). The typical scattering angle for an electron-plasmon interaction is around 13 µrad (assuming a 100 kV electron beam energy and a plasmon around 2 eV), this means that interference effects appear for angles < 13 µrad. Figure 6 shows a silver nanorod (aspect ratio of 6.6) in vacuum excited by a phase-shaped electron beam accelerated at 100 keV and using an entrance angle of 5 µrad. An initial Hermite-Gauss (first-order) wavefunction of the form $\Psi_i(\vec{R}) = xe^{-\frac{x^2+y^2}{w^2}}$ is used with w being the lateral extension of the beam (w=size of the rod). The configuration of lobes orientated along the rod long axis excites odd longitudinal modes ($L_1$ and $L_3$). On the other hand, by rotating the wavefunction by 90°, the lobes are orientated along to the rod short axis and even modes are preferentially exited ($L_2$ and $L_4$) and additionally the dipolar transverse mode (denoted as TM) is excited.

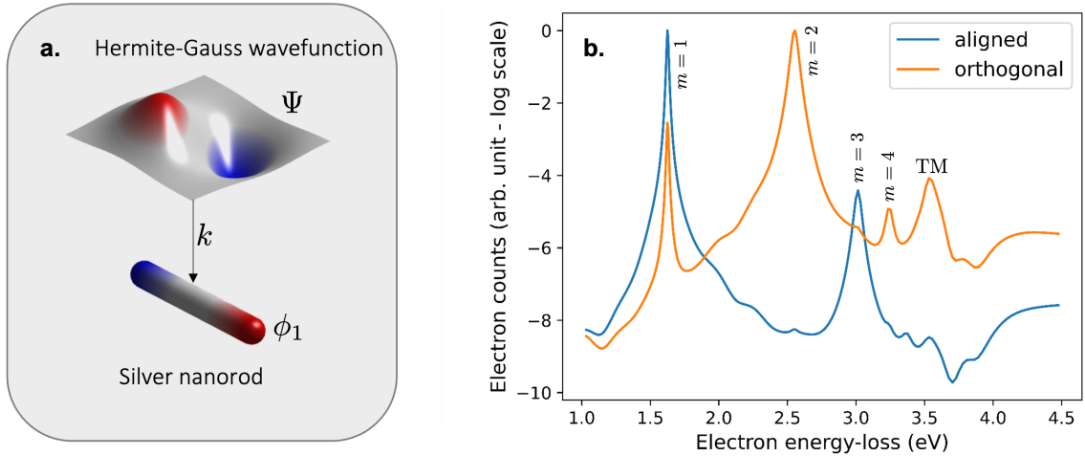

Figure 6: Selective probing of longitudinal and dipolar transverse modes through phase-shaped electron beam accelerated at 100 keV considering 5 µrad aperture entrance angle. (a) An initial Hermite-Gauss wavefunction impinges on a silver nanorod of aspect ratio 6.6 in vacuum. (b) EEL spectrum for lobes of the wavefunction aligned with the rod long axis direction (blue curve) and EEL spectrum for the orthogonal orientation of the wavefunction (orange curve). The dipolar transverse mode (denoted TM) is excited by the orthogonal orientation of the wavefunction.

# Conclusion

In summary, we used optical extinction and electron energy loss spectroscopy to study surface plasmon modes sustained by nanorods. Although dipolar longitudinal mode is well defined for both types of excitations, the dipolar transverse mode in EELS is hidden by overlapping with higher-order modes. Moreover, EELS reveals higher-order modes not observed in optical extinction (dark modes). An eigenmode analysis proved that for a low rod aspect ratio the influence of higher-order modes on the dipolar transverse mode decreases. However, the dipolar transverse mode is difficult to resolve even for a low aspect ratio rod due to the proximity in energy with the dipolar longitudinal mode. To explore more strategies to excite modes, a sample rotation with respect to the electron beam trajectory was performed and this improved the match in symmetry between the electron electric field and the transverse mode. By tilting a nanorod of aspect ratio equal to 2.73, there was a transition from second-order longitudinal mode to the dipolar transverse mode. It was necessary to rotate by the sample by 45° to resolve the transverse mode. These results allow us to optimize EELS experiments and obtain a more accurate interpretation of plasmon modes in nanorods. These observations can be extended to other geometries like disks and triangles to resolve in- and out-plane plasmon modes by EELS. In addition, by a theoretical description based on the wave mechanics, the EELS signal was decomposed in longitudinal and transverse phase-matching conditions which allowed us to understand the coupling between the electron beam and specific plasmon modes. In particular, the longitudinal phase-matching depends strongly on the electron energy revealing an optimization of the transverse mode signal by using a high-energy electron. The transverse phase-matching was studied using phase-shaped wavefunctions of Hermite-Gauss symmetry, simulating a pEELS situation. According to the alignment of the wavefunction with respect to a rod long axis, it was possible to selectively excite odd and even plasmon longitudinal modes. The transverse mode was obtained by using the wavefunction with lobes oriented parallelly to the rod short axis. The present work helps to understand plasmon excitations in nanorods by EELS and offers different novel strategies to overcome overlapping plasmon modes problems.

# Methods

Colloidal gold nanorods of small aspect ratios (< 2.5) were synthesized through the seedless mediated method.[47] Briefly, 5 mL of 1 mM $HAuCl_4$ (Sigma-Aldrich, EE.UU.), 5 ml of 200 mM CTAB (Sigma-Aldrich, Switzerland) and 270 µL of 4 mM $AgNO_3$ (Merck, Macedonia) were mixed. Then, 10 µL HCl at 37% (Scharlau, Spain) adjusted the pH close to one. Finally, 15 µL of 10 mM $NaBH_4$ (Merck, Germany) was added and the final solution was left at rest for 3 hours. In order to vary the aspect ratio, the volume of $AgNO_3$ was changed to 310 and 330 µL. The synthesized colloidal solutions were characterized by a UV-visible spectrophotometer (1800 UV-visible-Shimadzu, Japan) to confirm the rod plasmon signal. The aspect ratio was determined in a transmission electron microscope (JEOL 1010, USA Inc.). The standard deviation of the aspect ratio was around 0.5 for all synthesized samples. Higher aspect ratio rods (>2.5) were purchased from Sigma-Aldrich company. The standard deviation of the aspect ratio was around 0.2 for all commercial samples. The Solutions were drop-casted on (S)TEM $Si_3N_4$ membranes of 15 nm thickness. STEM-EELS measurements were performed in the spectral image (SI) mode in an UltraStem200 (Nion Co.) at 60 keV with ∼ 1 Å beam diameter. SI acquisition allows to get high-angle annular dark field (HAADF) signal and EEL spectrum simultaneously and compare pixel per pixel. Richardson-Lucy deconvolution[48] was performed (15 iterations) using Hyperspy,[49] achieving 185 meV of full width at half maximum (FWHM) of the zero-loss peak (ZLP). Monochromated STEM-EELS SI tilt series were collected in a Chromatem (Nion Co.) at 100 keV, 10 mrad EELS aperture and ∼ 1 Å beam diameter, achieving a ZLP FWHM of 26.8 meV. Simulations were performed by Boundary Element Method (BEM,[50]), and using the MNPBEM Matlab toolbox.[51] For pEELS simulation, equation (15) was used considering 30 modes, an aperture entrance angle of 5 µrad and the initial wavefunction dimension was identical to the nanorod size. The dielectric functions of gold and silver were taken from Johnson and Christy.[52]

## Acknowledgement

Thanks to project FID18-066 of SENACYT for financial support and CEMCIT-AIP for funding administration. Alfredo Campos thanks to Sistema Nacional de Investigación SNI de Panamá. Franck Aguilar thanks to project APY-GC-2019B-04, Senacyt, for internship funding. This project has received funding from the European Union's Horizon 2020 research and innovation programme under grant agreement No 823717 – ESTEEM3. We thank M. Sivis for making the render graphics of the rod in Figure S2 of the Supporting Information.

## Supporting Information Available

Details on experimental EEL spectra at different tilt angles in a gold nanorod are presented. Simulated EEL spectra at different electron energy in a gold nanorod are presented.

## Data Availability

The datasets generated and analyzed in this article are available from the corresponding author on reasonable request

## Author Contributions

All authors contributed to the study conception and design. Nanoparticle preparations were carried out by Damián Montero. STEM-EELS experiments were performed by Xiaoyan Li, Mathieu Kociak and Franck Aguilar. STEM-EELS data analysis was done by Franck Aguilar. Theoretical development was made by Hugo Lourenço-Martins and numerical simulations were performed by Hugo Lourenço-Martins and Alfredo Campos. The first draft of the manuscript was written by Franck Aguilar, Hugo Lourenço-Martins and Alfredo Campos and all authors commented on previous versions of the manuscript. All authors read and approved the final manuscript.

## Author Information

**Corresponding Author**

Alfredo Campos - Facultad de Ciencias y Tecnología, Universidad Tecnológica de Panamá, Panama, Panama, orcid.org/0000-0001-7921-8309; Email: alfredo.campos@utp.ac.pa

**Authors**

Franck Aguilar - Facultad de Ciencias y Tecnología, Universidad Tecnológica de Panamá, Panama, Panama; orcid.org/0000-0001-7950-5551

Hugo Lourenço-Martins - CEMES-CNRS, Université de Toulouse, CNRS, Toulouse, France; orcid.org/0000-0003-3929-5742

Damián Montero - Facultad de Ingeniería Mecánica, Universidad Tecnológica de Panamá, Panama, Panama;

Xiaoyan Li - Université Paris-Saclay, CNRS, Laboratoire de Physique des Solides, 91405 Orsay, France; orcid.org/0000-0003-2332-2523

Mathieu Kociak - Université Paris-Saclay, CNRS, Laboratoire de Physique des Solides, 91405 Orsay, France; orcid.org/0000-0001-8858-0449

Alfredo Campos - Facultad de Ciencias y Tecnología, Universidad Tecnológica de Panamá, Panama, Panama; orcid.org/0000-0001-7950-5551

**Notes** The authors declare no competing financial interest

## TOC Graphic

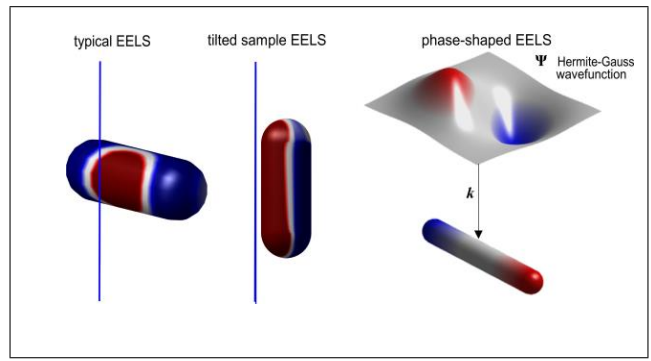